\documentclass[pra,prl,twocolumn,superscriptaddress,floatfix,showpacs,10pt,letterpaper]{revtex4}
\usepackage{graphicx}
\usepackage{subfigure}
\usepackage{amssymb}
\usepackage{verbatim}
\usepackage{amsmath}
\usepackage{amscd}
\usepackage{amssymb}

\newcommand{\ket}[1]{|{#1}\rangle}

\newcommand{\ncd}{\newcommand}
\ncd{\QC}{$\mbox{QC}_{\cal{C}}\;$}
\ncd{\QCpr}{${\mbox{QC}_{\cal{C}}}^\prime\;$}
\ncd{\QCns}{$\mbox{QC}_{\cal{C}}$}
\ncd{\QCprns}{${\mbox{QC}_{\cal{C}}}^\prime$}
\ncd{\cskN}{{|\phi_{\{\kappa\} } \rangle}_{{\cal{C}}_N}}
\ncd{\cskNpr}{{|\phi_{\{\kappa^\prime\} } \rangle}_{{\cal{C}}_N}}
\ncd{\cskNtil}{{|\phi_{\{\tilde{\kappa} \} } \rangle}_{{\cal{C}}_N}}
\ncd{\csk}{{|\phi_{\{\kappa\} } \rangle}_{\cal{C}}}
\ncd{\csktil}{{|\phi_{\{\tilde{\kappa} \} } \rangle}_{\cal{C}}}
\ncd{\cskf}{|\phi_{\{\kappa\} } \rangle_{\cal{C}}}
\ncd{\csktilf}{|\phi_{\{\tilde{\kappa} \} } \rangle_{\cal{C}}}
\ncd{\bracsk}{\mbox{}_{\cal{C}}\langle\phi_{\{\kappa\} }|}
\ncd{\bracsktil}{\mbox{}_{\cal{C}}\langle\phi_{\{\tilde{\kappa} \} }|}
\ncd{\nbracsk}{\mbox{}_{\cal{C}}\langle\phi_{\{\kappa\} }}
\ncd{\nbracsktil}{\mbox{}_{\cal{C}}\langle\phi_{\{\tilde{\kappa} \} }}
\ncd{\cs}{|\phi \rangle_{\cal{C}}\;}
\ncd{\csns}{|\phi \rangle_{\cal{C}}}
\ncd{\nbgh}{\text{nbgh}}
\ncd{\Sab}{S^{ab}}
\ncd{\Sba}{S^{ba}}
\ncd{\ds}{\displaystyle}
\ncd{\ovl}{\overline}

\newtheorem{fact}{Fact}

\newtheorem{example}{Example}

\newcommand{\nc}{\newcommand}
\nc{\rnc}{\renewcommand}
\nc{\beq}{\begin{equation}}
\nc{\eeq}{{\end{equation}}}
\nc{\beqa}{\begin{eqnarray}}
\nc{\eeqa}{\end{eqnarray}}
\nc{\lbar}[1]{\overline{#1}}
\nc{\ketbra}[2]{|#1\rangle\!\langle#2|}
\nc{\braket}[2]{\langle#1|#2\rangle}
\nc{\proj}[1]{| #1\rangle\!\langle #1 |}
\nc{\avg}[1]{\langle#1\rangle}
\nc{\Rank}{\operatorname{Rank}}
\nc{\smfrac}[2]{\mbox{$\frac{#1}{#2}$}}
\nc{\Tr}{\operatorname{Tr}}
\nc{\id}{\operatorname{id}}
\nc{\1}{\openone}
\nc{\ox}{\otimes}
\nc{\dg}{\dagger}
\nc{\dn}{\downarrow}
\nc{\cA}{{\cal A}}
\nc{\cB}{{\cal B}}
\nc{\cC}{{\cal C}}
\nc{\cD}{{\cal D}}
\nc{\cE}{{\cal E}}
\nc{\cF}{{\cal F}}
\nc{\cG}{{\cal G}}
\nc{\cH}{{\cal H}}
\nc{\cI}{{\cal I}}
\nc{\cJ}{{\cal J}}
\nc{\cK}{{\cal K}}
\nc{\cL}{{\cal L}}
\nc{\cM}{{\cal M}}
\nc{\cN}{{\cal N}}
\nc{\cO}{{\cal O}}
\nc{\cP}{{\cal P}}
\nc{\cR}{{\cal R}}
\nc{\cS}{{\cal S}}
\nc{\cT}{{\cal T}}
\nc{\cX}{{\cal X}}
\nc{\cY}{{\cal Y}}
\nc{\cZ}{{\cal Z}}
\nc{\var}{\operatorname{var}}
\nc{\rar}{\rightarrow}
\nc{\lrar}{\longrightarrow}
\nc{\polylog}{\operatorname{polylog}}

\nc{\RR}{{{\mathbb R}}}
\nc{\CC}{{{\mathbb C}}}
\nc{\FF}{{{\mathbb F}}}
\nc{\NN}{{{\mathbb N}}}
\nc{\ZZ}{{{\mathbb Z}}}
\nc{\PP}{{{\mathbb P}}}
\nc{\QQ}{{{\mathbb Q}}}
\nc{\UU}{{{\mathbb U}}}
\nc{\EE}{{{\mathbb E}}}
\nc{\Icoh}{{I^{\rm coh}}}
\nc{\Qca}{{Q_{\rm ss}}}
\nc{\Qcaa}{{Q^{(1)}_{\rm ss}}}
\nc{\Dcaa}{{D^{(1)}_{{\rm ss}\rightarrow}}}
\nc{\Dca}{{D_{{\rm ss}\rightarrow}}}

\nc{\be}{\begin{equation}}
\nc{\ee}{{\end{equation}}}
\nc{\bea}{\begin{eqnarray}}
\nc{\eea}{\end{eqnarray}}
\nc{\<}{\langle}
\rnc{\>}{\rangle}
\nc{\Hom}[2]{\mbox{Hom}(\CC^{#1},\CC^{#2})}
\nc{\rU}{\mbox{U}}

\begin{document}

\title{Generalized Concatenated Quantum Codes}

\author{Markus Grassl}
\affiliation{Institute for Quantum Optics and Quantum Information,
 Austrian Academy of Sciences, Technikerstra{\ss}e 21a, 6020
 Innsbruck, Austria} 

\author{Peter Shor}
\affiliation{Department of Mathematics, Massachusetts Institute of
Technology, Cambridge, MA 02139, USA}

\author{Graeme Smith}
\affiliation{IBM T.J. Watson Research Center, Yorktown Heights, NY 10598, USA}

\author{John Smolin}
\affiliation{IBM T.J. Watson Research Center, Yorktown Heights, NY 10598, USA}

\author{Bei Zeng}
\affiliation{IBM T.J. Watson Research Center, Yorktown Heights, NY 10598, USA}
\affiliation{Department of Physics, Massachusetts Institute of
Technology, Cambridge, MA 02139, USA}
\date{\today}

\begin{abstract}
We introduce the concept of generalized concatenated quantum
codes. This generalized concatenation method provides a systematical way for
constructing good quantum codes, both stabilizer codes and nonadditive codes. Using
this method, we construct families of new single-error-correcting
nonadditive quantum codes, in both binary and nonbinary cases, which
not only outperform any stabilizer codes for finite block length, but
also asymptotically achieve the quantum Hamming bound for large block
length.
\end{abstract}

\pacs{03.67.Lx} \maketitle 
Quantum error-correcting codes (QECCs) play a vital role in reliable
quantum information transmission as well as fault-tolerant quantum
computation (FTQC). So far, most good quantum codes constructed are
stabilizer codes, which correspond to classical additive codes. There
is a rich theory of stabilizer codes, and a thorough understanding of
their properties \cite{Gottesman, GF4}. However, these codes are
suboptimal in certain cases---there exist nonadditive codes which
encode a larger logical space than any stabilizer code of the same
length that is capable of tolerating the same number of errors
\cite{RHSS,SSW,Yu1}.

The recently introduced codeword stabilized (CWS) quantum codes
\cite{CWS1, CWS2, CWS3} framework, followed by the idea of union of
stabilizer codes construction \cite{Markus1, Markus2}, provides a
unifying way of constructing a large class of quantum codes, both
stabilizer codes and nonadditive codes. The CWS framework naturally allows to search
for good quantum codes, and some good nonadditive codes that
outperform any stabilizer codes have been found. However, this search
algorithm is very inefficient \cite{CWS2}, which prevents us from
searching for good quantum codes of length $n\geq 10$ in the binary
case and even smaller lengths in the nonbinary case. 

This letter introduces the concept of generalized concatenated quantum
codes (GCQCs), which is a systematical way of constructing good QECCs,
both stabilizer codes and nonadditive codes. Compared to the usual concatenated
quantum code construction, the role of the basis vectors of the inner
quantum code is taken on by subspaces of the inner code. The idea of
concatenated codes, originally described by Forney in a seminal book
in 1966 \cite{Forney}, was introduced to quantum computation community
three decades later \cite{Aharonov, Knill, Knill2, Zalka,
  Gottesman}. These concatenated quantum codes play a central role in
FTQC, as well as the study of constructing good degenerate QECCs.

The classical counterpart of GCQCs, i.e., generalized concatenated
codes, was introduced by Blokh and Zyablov \cite{Zyablov}, followed by
Zinoviev \cite{Zinoviev}. These codes improve the parameters of
conventional concatenated codes for short block lengths
\cite{Zinoviev} as well as their asymptotic performance
\cite{Zyablov2}. Many good codes, linear and nonlinear, can be
constructed from this method. One may expect that moving to the
quantum scenario, the GCQC method should be also a powerful one in
making good codes, which we show is the case.

We demonstrate the power of this new GCQC method by showing that some good 
stabilizer quantum codes, such as some quantum Hamming codes, can be constructed this way.
We then further construct families of nonadditive single-error-correcting CWS quantum codes, 
in both binary and nonbinary cases, which outperform any stabilizer
codes. This is the first known systematical construction of these good
nonadditive codes, while previous codes were found by exhaustive or
random numerical search with no structure to generalize to other cases.
We also show that these families of nonadditive codes asymptotically achieve the quantum
Hamming bound.

\textit{Basic Principle}
 A general quantum code $Q$ of $n$
$q$-dimensional systems, encoding $K$ levels, is a $K$-dimensional
subspace of the Hilbert space $\mathcal{H}_q^{\otimes n}$. We say $Q$
is of distance $d$ if all $d-1$ errors (i.e., operators acting
nontrivially on less than $d$ individual $\mathcal{H}_q$s) can be detected or
have no effect on $Q$, and we denote the parameters of $Q$ by
$((n,K,d))_q$. 

Recall that concatenated quantum codes are constructed from two
quantum codes, an \textit{outer} code $A$ and an \textit{inner} code
$B$. If $B$ is an $((n,K,d))_q$ code with basis vectors
$\{|\varphi_i\rangle\}_{i=0}^{K-1}$, then the outer code $A$ is taken
to be an $((n',K',d'))_K$ code, i.e., a subspace $A\subset
\mathcal{H}_K^{\otimes n'}$.  The concatenated code $Q_c$ is
constructed in the following way: for any codeword
$|\phi\rangle=\sum_{i_1\ldots i_{n'}}\alpha_{i_1\ldots i_{n'}}|i_1\ldots i_{n'}\rangle$
in $A$, replace each basis vector $|i_j\rangle$ (where
$i_j=0,\ldots,K-1$ for $j=1,\ldots,{n'}$) by a basis vector
$|\varphi_{i_j}\rangle$ in $B$, i.e.,
\begin{equation}
|\phi\rangle\mapsto |\tilde{\phi}\rangle=\sum_{i_1\ldots i_{n'}}\alpha_{i_1\ldots i_{n'}}|\varphi_{i_1}\rangle\ldots|\varphi_{i_{n'}}\rangle,
\end{equation}
so the resulting code $Q_c$ is an $((nn',K',\delta))_q$ code, and the
distance $\delta$ of $Q_c$ is at least $dd'$, for examples, see
\cite{Knill,Gottesman}.

In its simplest version, a generalized concatenated quantum code is
also constructed from two quantum codes, an \textit{outer} code $A$
and an \textit{inner} code $B$ which is an $((n,K,d))_q$ code. The
inner code $B$ is further partitioned into $r$ mutually orthogonal
subcodes $\{B_i\}_{i=0}^{r-1}$, i.e.
\begin{equation}\label{eq:code_decomposition}
B=\bigoplus_{i=0}^{r-1} B_i,
\end{equation}
and each $B_i$ is an $((n,K_i,d_i))_q$ code, with basis vectors
$\{|\varphi_{i,j}\rangle\}_{j=0}^{K_i-1}$, and $i=0,\ldots,r-1$. 

Now choose the outer code $A$ to be an $((n',K',d'))_r$ quantum code
in the Hilbert space $\mathcal{H}_{r}^{\otimes n'}$. While for
concatenated quantum codes each basis state $\ket{i}$ of the space
$\mathcal{H}_r$ is replaced by a basis state $\ket{\varphi_i}$ of the
inner code, for a generalized concatenated quantum code $Q_{gc}$ the
basis state $\ket{i}$ is mapped to the subcode $B_i$ of the inner
code.  For simplicity we assume that all subcodes $B_i$ are of equal
dimension, i.e., $K_1=K_2=\ldots=K_r=R$.  Then the dimension of the
resulting code $Q_{gc}$ is $\mathcal{K}=K'R^{n'}$, i.e., for each of
the $n'$ coordinates of the outer code, the dimension $\mathcal{K}$ is increased by
the factor $R$.  For a codeword
$\ket{\phi}=\sum_{i_1\ldots i_{n'}}\alpha_{i_1\ldots i_{n'}}\ket{i_1\ldots i_{n'}}$
of the outer code and a basis state $\ket{j_1\ldots j_{n'}}$ (where
$j_l=0,\ldots,R-1$ for $l=1,\ldots,n'$) of the space
$\mathcal{H}_R^{\otimes n'}$, the encoding is given by the
following mapping:
\begin{equation}\label{eq:twolevel_concatenation}
\ket{\phi}\ket{j_1\ldots j_{n'}}\mapsto
\sum_{i_1\ldots i_{n'}}\alpha_{i_1\ldots i_{n'}}
\ket{\varphi_{i_1,j_1}}\ldots\ket{\varphi_{i_{n'},j_{n'}}}.
\end{equation}

Note that the special case when $R=1$ corresponds to concatenated
quantum codes.  The resulting code $Q_{gc}$ has parameters
$((nn',\mathcal{K},\delta))_q$ where the distance $\delta$ is at least
$\min\{dd',d_i\}$.  If some of the $K_i$s differ, the calculation of
the dimension is more involved.

\textit{CWS-GCQC} 
From now on we restrict ourselves in constructing
some special kind of quantum codes, namely, CWS codes. CWS codes
include all the stabilizer codes and many good nonadditive codes
\cite{CWS1}, so it is a large class of quantum codes. The advantage of
the CWS framework is that the problem of constructing quantum codes is
reduced to the construction of some classical codes correcting certain
error patterns induced by a graph. So the point of view of
constructing these codes could be fully classical. For simplicity we
only consider nondegenerate codes here.

A nondegenerate $((n,K,d))_q$ CWS codes $Q_{\text{CWS}}$ is fully
characterized by a graph $\mathcal{G}$ and a classical code
$\mathcal{C}$ \cite{CWS1,CWS2,CWS3}, and for simplicity we only
consider $q$ a prime power. For any graph $\mathcal{G}$ of $n$ vertices,
there exists a unique stabilizer code $((n,1,d_{\mathcal{G}}))$
defined by $\mathcal{G}$ (called the graph state of $\mathcal{G}$). We
call the distance $d_{\mathcal{G}}$ the graph distance of
$\mathcal{G}$. For constructing a nondegenerate CWS code, we require
that the distance of the code be $\leq d_{\mathcal{G}}$. Then any
quantum error $E$ acting on $Q_{\text{CWS}}$ can be transformed into a
classical error by a mapping $Cl_{\mathcal{G}}(E)$ whose image is an
$n$-bit string.  The nondegenerate code $Q_{\text{CWS}}$ detects the
error set $\mathcal{E}$ if and only if $\mathcal{C}$ detects
$Cl_\mathcal{G}(\mathcal{E})$ \cite{CWS1,CWS2,CWS3}.

We take the inner code $B$ to be an $((n,K,d))_q$ nondegenerate CWS
code, constructed by a graph $\mathcal{G}$ and a classical code
$\mathcal{B}$. Furthermore, we decompose $B$ as
$B=\bigoplus_{i=0}^{r-1} B_i$ such that each $B_i$ is an
$((n,K_i,d_i))_q$ CWS code constructed from $\mathcal{G}$.  The basis
vectors of each $B_i$ can be represented by classical codewords of a
code $\mathcal{B}_i=\{\mathbf{b}_{i,j}\}_{j=1}^{K_i}$. Then
consequently, the classical code $\mathcal{B}$ has a partition
$\mathcal{B}=\bigcup_{i=0}^{r-1} \mathcal{B}_i$.

Now we take the outer code $A$ to be an $((n',K',d'=1))_r$ code in the
Hilbert space $\mathcal{H}_{r}^{\otimes n}$, which is constructed from
a classical $(n',K',d_c)_r$ code $\mathcal{A}$ over an alphabet of
size $r$, of length $n'$, size $K'$, and distance $d_c$ in the
following way: the basis vector $|\psi_{i_1\ldots i_{n'}}\rangle$ of
$A$ is given by
\begin{equation}
|\psi_{i_1\ldots i_{n'}}\rangle=|i_1\ldots i_{n'}\rangle,\ \forall
(i_1\ldots i_{n'})\in \mathcal{A}^{n'}.
\end{equation}

Denote the generalized concatenated code obtained from $A$ and $B$ by
$Q_{gc}$. It is straightforward to see $Q_{gc}$ is also a CWS code,
where the corresponding graph is given by $n'$ disjoint copies of the
graph $\mathcal{G}$. The corresponding classical code
$\mathcal{C}_{gc}$ is a classical generalized concatenated code with
inner code $\mathcal{B}=\bigcup_{i=0}^{r-1} \mathcal{B}_i$ and outer
code $\mathcal{A}$. The minimum distance of $Q_{gc}$ is at
least $\min\{d,d_i,d_{\mathcal{G}}\}$. However, the following
statement provides an improved lower bound.

\textbf{Main Result}: The minimum distance of ${Q}_{gc}$ is given by
$\min\{dd_c,d_i,d_{\mathcal{G}}\}$. 

We will not give a technical detailed proof of this result
here. Instead, since the proof idea can be illustrated clearly with a
simple example, we will analyze such an example, which also
illustrates a systematical method of constructing good nonadditive
quantum codes that outperform the best stabilizer codes.

\textit{Good Nonadditive Codes} We start taking the subcode
$B_0$ of the inner code $B$ to be the well-known $((5,2,3))_2$
code, the shortest one-error-correcting quantum code. As a CWS code,
this code can be constructed by a pentagon graph as well as a
classical code $\mathcal{B}_0=\{00000,11111\}$.  Further details
can be found in \cite{CWS1}, here we just focus on the classical error
patterns given by the mapping $Cl_{\mathcal{G}}$.  Since the pentagon
has graph distance $3$, the CWS code $B_0$ has distance at least
$3$ if $\mathcal{B}_0$ detects up to two errors with the error patterns
induced by the pentagon. The induced error patterns are given by the
following strings
\begin{eqnarray}
&Z:&\{10000,01000,00100,00010,00001\},\nonumber\\
&X:&\{01001,10100,01010,00101,10010\},\nonumber\\
&Y:&\{11001,11100,01110,00111,10011\}.
\label{pattern}
\end{eqnarray}
It is straightforward to check that $\mathcal{B}_0$ indeed detects
two of these errors. 

The classical code $\mathcal{B}_0$ is linear, so we can choose
$15$ disjoint proper cosets, e.g., $\mathcal{B}_1=\{00001,11110\}$
and $\mathcal{B}_{15}=\{01111,10000\}$. Combining these classical
codes with the pentagon gives us the CWS codes $B_i$, each of
which is a $((5,2,3))_2$ quantum code.  The union
$\mathcal{B}=\bigcup_{i=0}^{15} \mathcal{B}_i$ of all cosets is a
classical $(5,32,1)_2$ code which consists of all $5$-bit strings.
Combining $\mathcal{B}$ with a pentagon gives us the CWS quantum inner
code $B$ which is a $((5,32,1))_2$ quantum code. It can be decomposed
as ${B}=\bigoplus_{i=0}^{15} {B}_i$.

For the outer code we take a quantum code $A$ which corresponds to a
classical code $\mathcal{A}=(3,16,3)_{16}$, i.e., a distance three
code over $GF(16)$ of length $3$. Hence the basis of $A$ is given by
$\ket{i_1i_2i_3}$ where $(i_1i_2i_3)$ is one of the $16$ codewords of
$\{\text{\texttt{000}},\text{\texttt{111}},\ldots,\text{\texttt{aaa}},\ldots,\text{\texttt{fff}}\}$
of $\mathcal{A}$.  Here we use the hexadecimal notation to denote the
$16$ symbols of the alphabet $GF(16)$.  

\begin{figure}[h!]
\centering
\includegraphics[width=2.5in,angle=0]{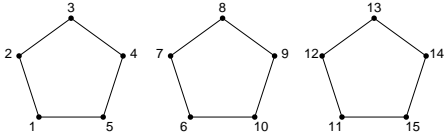}
\caption{Three pentagons: graph with $15$ vertices.}
\label{fig:3pentagons}
\end{figure}

Now we construct the GCQC ${Q}^{\{15\}}_{gc}$ of length $15$ from $A$
and $B$ in the following way: first, due to the product state form of
the basis of $A$, we choose the corresponding graph
$\mathcal{G}^{\{15\}}$ to be by three disjoint pentagons, as shown in
FIG. \ref{fig:3pentagons}. We denote this graph by
$\mathcal{G}^{\{15\}}$.  The distance of the graph state corresponding
to $\mathcal{G}^{\{15\}}$ is still $3$. So from these three pentagons
we can obtain a nondegenerate CWS quantum code whose distance is at
most $3$.  The error patterns induced by the mapping
$Cl_{\mathcal{G}^{\{15\}}}$ given by this $15$ vertex graph are simply
the strings from Eq. (\ref{pattern}) on the coordinates $1$--$5$ (or
$6$--$10$, or $11$--$15$) and zeros on the other coordinates. For
instance, $10000$ in Eq. (\ref{pattern}) gives rise to three strings
of length $15$ which are $100000000000000$, $000001000000000$, and
$000000000010000$. In total there are $45$ strings in the induced
error set of three pentagons corresponding to the $45$ single-qubit
errors on $15$ qubits.

Now we need to figure out what the corresponding classical code
$\mathcal{C}^{\{15\}}_{gc}$ is.  We know that it is the generalized
concatenated code with inner code $\mathcal{B}=\bigcup_{i=0}^{15}
\mathcal{B}_i$ and outer code $\mathcal{A}$. To see how this works
explicitly, consider the first codeword $\mathbf{a}_0=000$ of
$\mathcal{A}$.  Each of the three zeros is replaced by the code
$B_0=\{00000,11111\}$, i.\,e, $(\mathbf{a}_0,j_1,j_2,j_3)$ (where
$j_l=0,1$) will be mapped to one of the $8$ codewords of
$\mathcal{C}^{\{15\}}_{gc}$, which are strings of length $15$, given
by $000000000000000$, $000000000011111$, $000001111100000$,
$000001111111111$, $111110000000000$, $111110000011111$,
$111111111100000$, $111111111111111$.  Similarly, any other codeword
$\mathbf{a}_i$ of $\mathcal{A}$ will be mapped to $2^3$ codewords in
$\mathcal{C}^{\{15\}}_{gc}$ obtained by concatenating three codewords
of $B_i$. The size of $\mathcal{C}^{\{15\}}_{gc}$ is then
$2^3\times16=2^7$.

We now show that the distance of ${Q}^{\{15\}}_{gc}$ is $3$. To see
this, we only need to show that $\mathcal{C}^{\{15\}}_{gc}$ detects up
to two errors of the error patterns induced by three pentagons. This
is clear via the following two observation: i)
$\mathbf{c}_1,\mathbf{c}_2\in\mathcal{C}^{\{15\}}_{gc}$ correspond to
different codewords of the outer code $\mathcal{A}$: since the
pentagons are disjoint, and $\mathcal{A}$ has distance $3$, at least
$3$ strings in the induced error patterns are needed to transform
$\mathbf{c}_1$ to $\mathbf{c}_2$. ii)
$\mathbf{c}_1,\mathbf{c}_2\in\mathcal{C}^{\{15\}}_{gc}$ correspond to
same codewords of the outer code $\mathcal{A}$: since at least $3$
strings in the induced error patterns are needed to transform codewords in
$\mathcal{B}_i$, at least $3$ strings in the induced error are needed
to transform $\mathbf{c}_1$ to $\mathbf{c}_2$.

Now one can generalize the construction of ${Q}^{\{15\}}_{gc}$ to the
case of more than three pentagons. Suppose we use $n'$ pentagons to
construct single-error-correcting CWS codes, then we observe the
following
\begin{fact}
Choose the inner code as ${B}=\bigcup_{i=0}^{15} {B}_i$ with each
$B_i$ a $((5,2,3))_2$ quantum code, and the outer code $A$
corresponding to the classical code $\mathcal{A}$ with parameters
$(n',K',3)_{16}$, then the resulting GCQC $Q_{gc}$ is a
$((5n',2^{n'}K',3))_2$ binary quantum code.
\end{fact}

This indicates that if we have a good classical code over $GF(16)$ of
distance $3$, then we may systematically construct good quantum codes
via the generalized concatenation method described above.
\begin{example}
Using the quantum code corresponding to the
classical Hamming code with parameters $(17,16^{15},3)_{16}$
as the outer code, then by \textbf{Fact 1} we get a quantum code with
parameters $((85,2^{77},3))_2$, which is a quantum Hamming code
\cite{GF4}. If we properly choose the labeling of the subcodes $B_i$
by elements of $GF(16)$, the correponding classical code is linear
\cite{Dumer}, and hence this quantum code is a stabilizer code \cite{CWS1}.
\end{example}

If we take a quantum code corresponding to a 
good nonlinear classical code as the outer code, then we 
can construct a good nonadditive quantum code \cite{CWS1}. 
Here we give examples of such a good quantum codes which are
constructed using a good nonlinear classical codes. Those nonlinear codes
are obtained via the following classical construction, called `subcode
over subalphabet' (see \cite[Lemma 3.1]{Dumer}).
\begin{fact} 
If there exists an $(n,K,d)_q$ code, then for any
$s<q$, there exists an $(n',K',d)_s$ code with size at least
$K(s/q)^n$.
\end{fact}
\begin{example}
It is known that there is a classical Hamming code with parameters
$(18,17^{16},3)_{17}$. Therefore, using \textbf{Fact 2} there is a $(18,\lceil
\frac{16^{18}}{17^2}\rceil,3)_{16}$ code. Then the resulting quantum code
has parameters $((90,2^{81.825},3))_2$. For a binary quantum code with
$n=90$ and $d=3$, the quantum Hamming bound ($K\leq q^n/((q^2-1)n+1)$, see \cite{GF4})
gives $K<2^{81.918}$, and the linear programing bound (see \cite{GF4}) gives
$K<2^{81.879}$. So the best stabilizer quantum code can only be
$((90,2^{81},3))_2$. Hence our simple construction gives a nonadditive
single-error-correcting quantum code which outperforms any possible stabilizer codes. This is
the first such example given by construction, not by numerical search.
\end{example}

\begin{example}
The similar CWS-GCQC idea works also for the nonbinary case using the
nonbinary CWS construction \cite{CWS3}. 
Take the inner code to be a union of $81$ mutually
orthogonal $((10,729,3))_3$ codes that is constructed from a graph
that is a ring of ten vertices \cite{Looi}. Choose the outer code as
the quantum code corresponding to the classical $(84,\lceil
\frac{81^{84}}{83^2}\rceil,3)_{81}$, which is obtained from the
Hamming code $(84,83^{82},3)_{83}$. Then the resulting quantum code
has parameters $((840,3^{831.955},3))_3$. For a ternary quantum code
with $n=840$ and $d=3$, the Hamming bound gives $K<3^{831.978}$, and
the linear programing bound gives $K<3^{831.976}$, so the best
stabilizer code can only be $((840,3^{831},3))_3$. This is the first known 
nonbinary nonadditive code which outperforms any stabilizer codes.
\end{example}

It is straightforward to generalize the above construction for binary
and ternary codes to build good nonadditive quantum codes in Hilbert
space $H_q^{\otimes n}$ for any prime power $q$.  For this, we take
the inner code $B_0$ as the perfect quantum Hamming code
$((q^{n_s},q^{n_s-2s},3))_q$ in $H_q^{\otimes n}$ of length
$n_s=(q^{2s}-1)/(q^2-1)$.  The full space $B=((n_s,q^{n_s},1))_q$ can be
decomposed as the sum of $q^{2s}$ orthogonal translates of $B_0$. The
outer quantum code is then corresponding to a classical code over an
alphabet of size $Q=q^{2s}$ given by \textbf{Fact 2}, i.e., the classical code is
obtained from the $P$-ary Hamming code $[L_i,L_i-i,3]_P$ where $P$ is
the least prime power exceeding $Q$, and $L_i=(P^i-1)/(P-1)$.  The
result is the code $V_{si}=((N_{si},M_{si},3))_q$ with length
$N_{si}=L_in_s=(P^i-1)(Q-1)/(q^2-1)(P-1)$ and dimension $M_{si}\geq
q^{N_{si}}/P^i$.

The number of different errors we want to deal with is
$(q^2-1)N_{si}+1>Q^i=q^{si}$ for $P>Q$ and $i>1$.  By the quantum
Hamming bound $K\leq q^{N_{si}}/((q^2-1)N_{si}+1)<q^{N_{si}}/Q^i$, the
dimension of any stabilizer code (including degenerate codes) is upper
bounded by $K\le q^{N_{si}-2si-1}$.  Hence for any prime power $P$
with $Q^i < P^i < q Q^i$, the dimension $M_{si}$ is strictly larger
than $q^{N_{si}-2si-1}$, i.e., our codes are better than any stabilizer
codes.  Moreover, we have $q^{N_{si}}/P^i \le M_{si}\le
q^{N_{si}}/Q^i$.  Since $Q/P\rightarrow 1$ for $s\rightarrow\infty$
\cite{Dumer}, these families of nonadditive codes asymptotically
achieve the quantum Hamming bound.

\textit{Discussion\/} We have introduced the concept of GCQC, which is
a systematic construction of good QECCs, both stabilizer codes and nonadditive codes.
One way of generalizing the concatenation of
Eq. (\ref{eq:twolevel_concatenation}) is 
to put some constraints on
the additional degrees of freedom $\ket{j_1\ldots j_{n'}}$ by using a
second outer code. Additionally, one can 
recursively decompose the
codes $B_i$ in the decomposition (\ref{eq:code_decomposition}) of the
inner code, which leads to a more general construction of GCQC 
with which more good quantum codes can be constructed (see \cite{MSZ}).
While the nonadditive codes of this letter tighten the gap between
lower and upper bounds for the dimension of the codes, 
we believe that
in general the GCQC construction gives a promising way for further
constructing new quantum codes of good performance, and we
hope that this generalized concatenation technique will also shed
light on improvements of fault-tolerant protocols.



\begin{thebibliography}{22}
\bibitem{Gottesman} D. Gottesman, Ph.D. Thesis, Caltech, 1997. arXiv: quant-ph/9705052.

\bibitem{GF4} A. R. Calderbank, E. M. Rains, P. W. Shor,
 N. J. A. Sloane, IEEE Trans. Inf. Theory, \textbf{44}, 1369 (1998).

\bibitem{RHSS} E. M. Rains, R. H. Hardin, P. W. Shor, and N. J. A. Sloane,  Phys. Rev. Lett.
\textbf{79}, 953 (1997).

\bibitem{SSW} J. A. Smolin, G. Smith, and S. Wehner, Phys. Rev. Lett. \textbf{99}, 130505 (2007).

\bibitem{Yu1} S. Yu, Q. Chen, C. H. Lai, and C. H. Oh,
 Phys. Rev. Lett. \textbf{101}, 090501, (2008).

\bibitem{CWS1} A. Cross, G. Smith, J. Smolin, and B. Zeng, IEEE Trans. Inf. Theory, \textbf{55}, 433 (2009).

\bibitem{CWS2} I. Chuang, A. Cross, G. Smith, J. Smolin, and B. Zeng, arXiv: 0803.3232.

\bibitem{CWS3} X. Chen, B. Zeng, and I. Chuang, Phys. Rev. \textbf{A78}, 062315 (2008).

\bibitem{Markus1} M. Grassl and M. R{\"o}tteler, Proc. 2008 IEEE
 Int. Symp. Inform. Theory, pp. 300--304 (2008). arXiv: 0801.2150.

\bibitem{Markus2} M. Grassl and M. R{\"o}tteler, Proc. 2008 IEEE
 Inf. Theory Workshop, pp. 396--400 (2008).  arXiv: 0801.2144.

\bibitem{Forney} G. D. Forney, Jr. \textit{Concatenated Codes}, Cambridge, MA: M.I.T. Press, 1966.

\bibitem{Knill} E. Knill, and R. Laflamme, arXiv: quant-ph/9608012.

\bibitem{Knill2} E. Knill, R. Laflamme, and W. Zurek, arXiv: quant-ph/9610011 (1996); E. Knill, R. Laflamme, and W. Zurek, arXiv: quant-ph/9702058 (1997). 

\bibitem{Zalka} C. Zalka, arXiv: quant-ph/9612028 (1996). 

\bibitem{Aharonov} D. Aharonov and M. Ben-Or, Prof. 29th Ann. ACM Symposium
 on Theory of Computing, pp. 176-188 (1997).  arXiv: quant-ph/9611025. 

\bibitem{Zyablov} E. L. Blokh and V. V. Zyablov, Probl.  Peredachi Inform. \textbf{10}, 45 (1974).

\bibitem{Zinoviev} V. A. Zinoviev,  Probl. Peredachi Inform., \textbf{12}, 5 (1976).

\bibitem{Zyablov2} E. L. Blokh and V. V. Zyablov, \textit{Linear Concatenated Codes}, Moscow: Nauka, 1982 (in Russian).

\bibitem{Dumer} I. Dumer, \textit{Concatenated Codes and Their Multilevel Generalizations}, Chapter 23, pp. 1911--1988. 
In \textit{Handbook of Coding Theory}, V. S. Pless and W. C. Huffman (eds.), Elsevier Science, Amsterdam (1998).

\bibitem{Looi} S. Y. Looi, L. Yu, V. Gheorghiu, and R. B. Griffiths, Phys. Rev. \textbf{A78}, 042303 (2008).

\bibitem{MSZ} M. Grassl, P. W. Shor, and B. Zeng, in preparation.

\end{thebibliography}
\end{document}